  \documentclass[aps,prd,twocolumn,superscriptaddress,floatfix]{revtex4-2}
 \usepackage{appendix}
\usepackage{color}
\usepackage{amssymb}
\usepackage[normalem]{ulem}
\usepackage{graphicx,color}
\usepackage{amsmath}
\usepackage{subfigure}
\usepackage[normalem]{ulem}
\usepackage{booktabs}
\usepackage{xcolor}
\usepackage{epsfig,graphics,color,graphicx,amsmath}
\usepackage{axodraw2}
\usepackage{enumitem}
\colorlet{darkgreen}{green!50!black}
\colorlet{brightyellow}{yellow!75!red}
\colorlet{orange}{red!50!yellow}
\colorlet{darkblue}{blue!60!black}
\colorlet{darkred}{red!80!black}

\usepackage{soul}
\newcommand{\pslash}{\slash \hspace{-0.22cm} p}
\newcommand{\kslash}{\slash \hspace{-0.22cm} k}

\usepackage{mathtools}
\usepackage{mathrsfs}
\usepackage{bm}
\usepackage{relsize}
\usepackage{indentfirst}

\usepackage{xcolor}

\usepackage{amssymb,amsmath,multirow,epsfig,graphicx,color}

\usepackage{epsfig}
\usepackage{graphicx,amsmath}
\def\be{\begin{eqnarray} &&}

	\def\ee{\end{eqnarray}}

\newcommand\ba{\begin{eqnarray}}
	\newcommand\ea{\end{eqnarray}}

\newcommand{\bas}{\begin{eqnarray*}}
	\newcommand{\eas}{\end{eqnarray*}}

\newcommand{\bno}{\begin{eqnarray*}}
	\newcommand{\eno}{\end{eqnarray*}}

\def\sl
\usepackage{hyperref}
\usepackage{url}
\usepackage{hyperref}
\hypersetup{
	colorlinks=true,   %% false,  
	linkcolor=black,   %% red, blue,
	filecolor=magenta,      
	urlcolor=blue,     %% black,    %% cyan,
	citecolor=black,    %% Red   %% blue   %% black
} 
\begin{document}

\hspace{1.75cm} \ {\bf \bf LFTC - 22 - 4/72}

\title{ Pion model with the Nakanishi Integral Representations  }
%Minkowski space pion model inspired  by  lattice  QCD  running  quark  mass,
\author{R.~M.~Moita }
\affiliation{Instituto Tecnol\'ogico de Aeron\'autica,  DCTA,
12228-900 S\~ao Jos\'e dos Campos,~Brazil}
\author{J.~P.~B.~C.~de Melo}
%\email[E-mail: ]{joao.mello@cruzeirodosul.edu.br}
\affiliation{Laborat\'orio de F\'\i sica Te\'orica e 
Computacional - LFTC, 
\\
Universidade Cruzeiro do Sul and 
Universidade Cidade de S\~ao Paulo (UNICID) 
\\  01506-000 S\~ao Paulo, Brazil}
\author{T.~Frederico}
\affiliation{Instituto Tecnol\'ogico de Aeron\'autica,  DCTA,
12228-900 S\~ao Jos\'e dos Campos,~Brazil}
\author{W.~de Paula}
\affiliation{Instituto Tecnol\'ogico de Aeron\'autica,  DCTA,
12228-900 S\~ao Jos\'e dos Campos,~Brazil}

\begin{abstract}
\vspace{1em}
In the present work, we describe a model for the pion based on an analytic expression for the Bethe-Salpeter (BSA)
amplitude, combined with some
ingredients from Lattice QCD calculations.
The running  quark mass function $M (p^2)$, used here, reproduces  well the results of Lattice QCD calculations.
The analytical form  of the running quark mass function contains a single time-like pole, which implies in 
time-like poles of the dressed quark propagator. Such a form allows to build the weight functions, $G_i(\gamma,z)$, for the
Nakanishi integral representation  of each scalar function, $\chi_{i}(k,p)$, appearing in the decomposition 
of the Bethe-Salpeter amplitude in terms of Dirac operators, Such scalar amplitudes can also be used to obtain 
the pion valence light-front wave function.
\vspace{1em}
\end{abstract}

%%\begin{keysword}{QCD, Pion, Bethe-Salpeter amplitude,  Nakanishi Integral Representation}
%% \vspace{-0.52cm}
%% \pacs{11.10.−z, 12.38.−t, 12.39.−x}
%\begin{keyword} 
%Pseudoscalar Meson, Nakanishi, Decay Constant, 
%Constituent Quark Model
%\end{keyword}
%\end{frontmatter}
\date{\today}
\maketitle

%%% \begin{multicols}{2}

Nowadays, the pion is understood  as a pseudo-scalar bound state of constituents 
carrying the fundamental degrees of freedom of 
the strong interaction theory
and, due to its  small mass at the hadronic scale, it is considered  a
 Goldstone boson~\cite{Zuber1980}. 
The special nature of the pion is associated with the spontaneous breaking of chiral symmetry, 
where the  light quarks acquires, dynamically,  sizable masses departing 
from their small current quark masses  due
to weak Higgs coupling.  A trace of that is found in  the small 
pion mass (0.140~GeV), which  would be zero
for vanishing current quark masses when  the chiral symmetry is exact.
Therefore, the pion acquires a mass by the explicit breaking of this symmetry, and 
it is the Goldstone boson associated with the Dynamical Chiral Symmetry Breaking phenomena~(DCSB),
which is well established within the theory of strong interactions, namely
Quantum Chromodynamics (QCD)~\cite{Cloet2014}. While the current masses of the
light quarks are small, the heavy ones have their large masses basically  due
to the Higgs coupling, breaking strongly the flavor symmetry, which was explored 
in a recent study  of  the flavor content of the light and  heavy pseudoscalar mesons~\cite{Moita2021}.  

In the present work, we will use the results from QCD calculations in the 
Landau gauge on the Euclidean Lattice~\cite{Parappilly2005} 
for the dressed light quarks   running masses,  as proposed in~\cite{Clayton2017} 
to model the quark propagator and the pion Bethe-Salpeter amplitude. Our aim is 
to explore the Nakanishi integral representation  of the pion Bethe-Salpeter 
amplitude by computing  each weight function, $G_i(\gamma,z)$, associated with the 
four scalar functions, $\chi_{i}(k,p)$, found in the decomposition  of the pion  
Bethe-Salpeter amplitude in Dirac spinorial space. 

The  general form of the dressed quark propagator is given by:
\begin{equation}
\label{sfO}
S_F(k)=\imath\,Z(k^2)\left[\kslash-M(k^2)+\imath\epsilon\right]^{-1} 
\ ,
\end{equation}
for the light quarks, namely, $u$ and $d$.  The dressed quark mass function
is $M(k^2)$, which is chosen to reproduce the results obtained from Euclidean Lattice QCD (LQCD) 
calculations~\cite{Parappilly2005}.
The quark wave function renormalization factor is taken here  as  $Z(k^2)=1$, 
for simplification  of the model~\cite{Clayton2017}, while it still captures 
the main physics of the  QCD  dynamical chiral  symmetry breaking  brought by 
the running  dressed quark mass function.

The model dressed quark propagator is given by:
\begin{equation}
S_F(k)=\imath 
\frac{\kslash + [ m_0 -m^3 (k^2-\lambda^2 + \imath \epsilon)^{-1}] }
{\left( k^2 -  ([ m_0 -m^3 (k^2-\lambda^2 + \imath \epsilon)^{-1})^2  ] + 
\imath\epsilon \right)}~,
\label{propagador}
\end{equation}
in which we can identify the running quark dressed mass function:
\begin{equation}
\label{runningmass}
M(k^2)=m_0-m^3\left[k^2- \lambda^2 +i \epsilon \right]^{-1}\, ,
\end{equation}
where 
$m_0=0.014 \,\text{GeV},\,\, m=0.574\,\text{GeV}\,\, \text{and}
\,\,\lambda\,=\,0.846\,\text{GeV}
$. For convenience,  we call this set as input parameters (IP)~\cite{Clayton2017,Moita2019}. 

They are chosen to fit the dressed light quark mass from 
LQCD~\cite{Parappilly2005} (see also~\cite{Rojas2013,Oliveira:2018ukh}) 
for space-like momenta as reproduced in the left panel of  Fig.~[\ref{massrunning}].

 The  dressed quark propagator in the present model has time-like  poles,
 found by  solving $m^2_i = M^2(m^2_i)$,
which allows to  write it in a factorized form:
\begin{equation}
S_F(k) =  \imath\,\,\frac{\left(k^2- \lambda^2\right)^2\, 
	(\kslash +m_0)- \left(k^2- \lambda^2\right)\,
	m^3}
{{\prod_{i=1,3}}(k^2-m^2_i+\imath \epsilon)}. 
\label{sf1}
\end{equation}
With the set (IP), we have the following poles masses,
$m_1=$  0.371~GeV, $m_2=$ 0.644~GeV and $m_3=$ 0.954~GeV~\cite{Clayton2017}.

%% \end{multicols}

\begin{widetext}

\vspace{0.8cm}
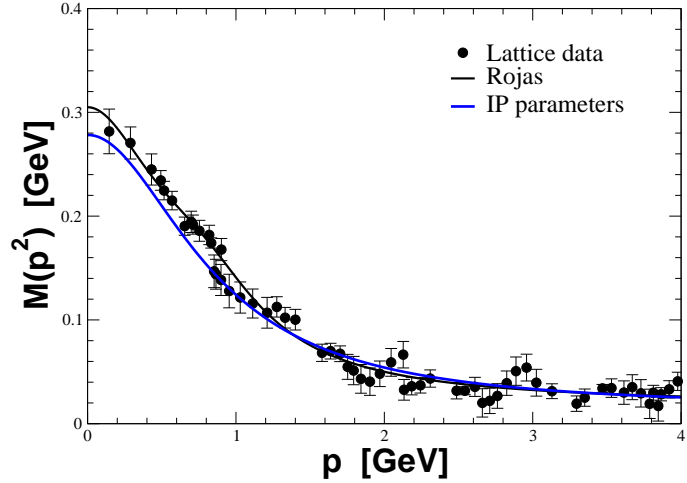
\begin{figure}[htb]
	%% \begin{center}
	 \epsfig{figure=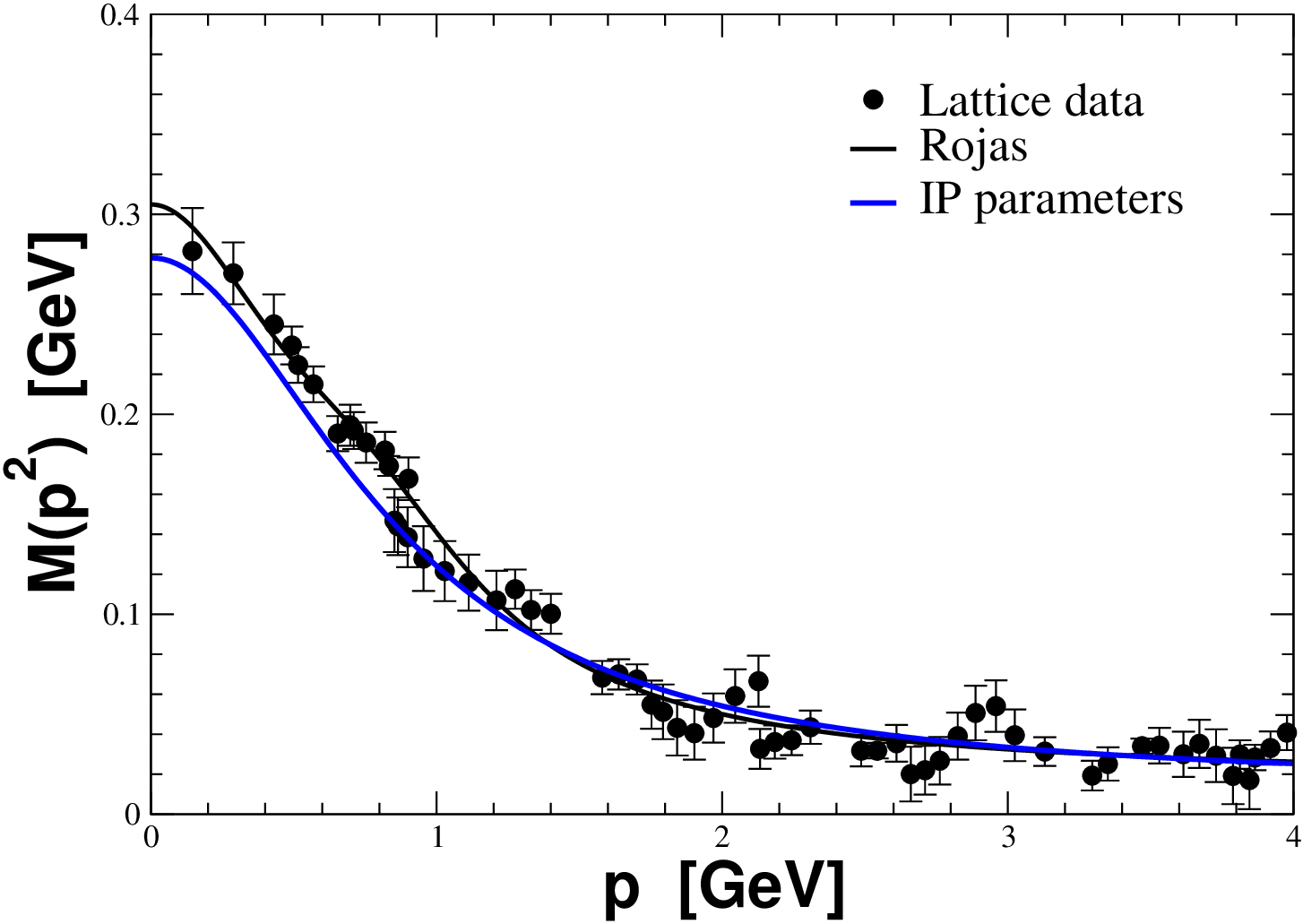,width=9.0cm}
 %% 	\hspace{-1.600220003cm}
 %%% 	\vspace{1.5cm}
 \begin{LARGE}%%Large}
%% \begin{figure}[h]
 %%\begin{center} 
% \vskip 1.cm
%%%  \SetScale{0.8}
\begin{picture}(330,130)(0,0)
 \Line(120,52)(150,52)
 \Line(135,55)(140,50)
 \Line(135,45)(140,50)
 \Line(120,48)(150,48)
 \Vertex(150,50){3.0}
 \put(155,25){\makebox(0,0)[br]{$\Gamma_\pi$}}   
 \ArrowLine(170,70)(150,50)
 \ArrowLine(150,50)(170,30)
 \put(218,12){\makebox(0,0)[br]{$k+p/2$}}  
 \put(218,70){\makebox(0,0)[br]{$k-p/2$}}  
 \put(110,45){\makebox(0,0)[br]{$p$}}
  \put(290,45){\makebox(0,0)[br]{$= \varPsi_{\pi}(k,p)$}}  
\end{picture}
\end{LARGE} 
	\caption{Top panel: dressed  quark running mass for the present model in the space-like momentum region, 
		compared with LQCD results in the Landau gauge~\cite{Parappilly2005}, and the parametrization 
		from Rojas et al.~\cite{Rojas2013}. Bottom panel: diagrammatic representation of the Bethe-Salpeter Amplitude.} 
	\label{massrunning}
	%% \end{center}
\end{figure}

\end{widetext}

%%   \begin{multicols}{2}
  %% 
The dressed quark propagator can be written as following
\begin{equation}
\label{sfaltern}
S_F(k)=\imath\,\left[A(k^2)\,\kslash+B(k^2)\right] \, , 
	\end{equation}
which by comparison with Eq.(\ref{sf1}), one gets the explicit expressions for  $A(k^2)$ and $B(k^2)$, as:
\begin{eqnarray}
A(k^2) & = &  \dfrac{\left(k^2- \lambda^2\right)^2}{{\prod_{i=1,3}}(k^2-m^2_i+\imath \epsilon)} \, ,
\nonumber \\
B(k^2)& = & 
\dfrac{(\lambda^2-k^2)m^3}{{\prod_{i=1,3}}(k^2-m^2_i+\imath \epsilon)}\,+\,m_0\,A(k^2)\, .
\label{defAB}
\end{eqnarray}

We can decompose $A(k^2)$ and $B(k^2)$ in the form of polynomials as:
\begin{eqnarray} 
A(k^2) & = &  \sum_{i=1}^{3} \frac{D_i}{k^2-m_i^2}
\nonumber \\
\,\,\text{and}\,\,
B(k^2) & = &  \-\-\-\- \sum_{i=1}^{3} \frac{m_0D_i -E_i}{k^2-m_i^2}\, ,
\label{set1}
\end{eqnarray}
where the residue are obtained from the set (IP):
\begin{eqnarray*}
&&D_1=1.4992, \, D_2=-0.5941, \,  D_3=-0.09498,
\nonumber \\
&& 
E_1=0.4240, \, E_2=-0.3314, \,
E_3=-0.07864 \, ,
\label{valorDE}
\end{eqnarray*}
with $D_i$ dimensionless and $E_i$ in units of GeV.

For our purpose, we can also describe the functions $A(k^2)$ and $B(k^2)$ 
in terms of a spectral representation:
\begin{eqnarray} 
\label{specrep}
A(k^2) & = &  \int_0^\infty d\mu^2 
\frac{\rho_A(\mu^2)}{k^2-\mu^2+\imath\varepsilon}, \ 
\nonumber \\ 
B(k^2) &  = & \int_0^\infty d\mu^2 
\frac{\rho_B(\mu^2)}{k^2-\mu^2+\imath\varepsilon} \ , 
\end{eqnarray}
where the spectral densities are:
$$
\rho_A(\mu^2)=-\frac1\pi \, \text{Im\,}[A(\mu^2)]\,\, \,\,
\text{and}\,\,\,\,\rho_B(\mu^2)=-\frac1\pi \, \text{Im\,}[B(\mu^2)]
~.
$$ 
One can easily check that the model spectral functions violate the positivity constraints~\cite{Zuber1980}:
$$ \label{posab}
\mathcal{P}_a=\rho_A(\mu^2)\geq 0 \,\, \,\,\text{and}
\,\,\,\,\mathcal{P}_b=\mu\,\rho_A(\mu^2)-\rho_B(\mu^2)\geq 0 \, , $$
which is not a problem as the quark cannot be an asymptotic state, as it should be confined within the hadron.

Remembering that we can write the functions $A(k^2)$ and $B(k^2)$ as a sum 
of polynomials, and combining with the spectral representation,
\begin{eqnarray}
\int_0^\infty d\mu^2 \frac{\rho_A(\mu^2)}{k^2-\mu^2+i\epsilon}
= \nonumber \\
\sum_{i=1}^3 \int_0^\infty d\mu^2
\frac{D_i~\delta(\mu^2-m_i^2)}{k^2-\mu^2+i\epsilon} & & 
,
\end{eqnarray}
we find the spectral density: 
$$
\rho_A(\mu^2)=D_1\,\delta(\mu^2-m_1^2)+D_2\,
\delta(\mu^2-m_2^2)+D_3\,\delta(\mu^2-m_3^2).
$$
For $B(k^2)$, the spectral decomposition is given by:
\begin{eqnarray}
\label{rhoB0}
\int_0^\infty d\mu^2 \frac{\rho_B(\mu^2)}{k^2-\mu^2+i\epsilon} =  
\nonumber \\ 
 \sum_{i=1}^3 \int_0^\infty 
d\mu^2\frac{E_i~\delta(\mu^2-m_i^2)}{k^2-\mu^2+i\epsilon}.
& & 
\end{eqnarray}
We obtain, for $\rho_B$, the following final expression
\begin{eqnarray*}
\label{rhoB}
\rho_B(\mu^2) & = & E_1\,\delta(\mu^2-m_1^2)+E_2\,\delta(\mu^2-m_2^2)
\\
& + & E_3\,\delta(\mu^2-m_3^2)+m_0\,\rho_A(\mu^2) 
.
\end{eqnarray*} 

In the present work, we use the Nakanishi Integral Representation~(NIR),
(see in ~\cite{dePaula:2016oct, dePaula:2017ikc} for more references), 
in order to write the Bethe-Salpeter amplitude for the pion quark-antiquark bound state. 
The first step is to write the pion-quark-antiquark vertex,
denoted by $\Gamma_\pi(k,p)$, which composes the pion Bethe-Salpeter amplitude,
diagrammatically represented in the right panel of Fig.~\ref{massrunning}. The most general form is given by:
\begin{eqnarray}
\label{vertex}
\Gamma_\pi (k,p) =  \gamma_5 [\imath E_\pi (k,p)+ \pslash  F_\pi (k,p)\qquad\qquad  
\nonumber \\
 +   k^\mu p_\mu \ \kslash G_\pi (k,p) +  \sigma_{\mu\nu} k^\mu p^\nu H_\pi (k,p)] \, .
\end{eqnarray}
The pion Bethe-Salpeter amplitude  has the form:
\begin{equation}
\Psi_\pi (k,p)  = S_F(k+\tfrac{p}{2})\,\Gamma_\pi(k,p)\,S_F(k-\tfrac{p}{2}), 
\label{BSA}
\end{equation}
with the vertex function~\cite{Clayton2017}
\begin{eqnarray}
\Gamma_\pi (k,p)  =    \imath \,\mathcal{N}\,\gamma_5 \,M (k)|_{m_0=0}
 & &   \nonumber \\
= -\imath \frac{\mathcal{N} \gamma_{5} m^{3}}{k^{2}
-\lambda^{2}+\imath \epsilon}
& & 
\, ,
\end{eqnarray}
dominated by the dressed quark mass function in the chiral limit. $\mathcal N$ is a normalization factor.

After defining the structure of the pion vertex,
we can write its BS amplitude, incorporating the dressed quark propagator, which also carries DCSB effects.
Using the compact notation
 for the propagators, one has that:
\begin{eqnarray}
\Psi_{\pi}(k,p)  & = &   -\left[A\left(k_{q}^{2}\right)
\kslash_{q}+B\left(k_{q}^{2}\right) \right]
 \frac{\mathcal{N} \gamma_5 m^3}{k^2-\lambda^2+\imath \epsilon}
\nonumber \\  
& & \times \left[A \left(k_{\bar{q}}^{2}\right)
\kslash_{\bar{q}}+B\left(k_{\bar{q}}^{2}
\right) \right]\, ,
\end{eqnarray}
here the quark and antiquark momentum are: $k_{q}=(k+p / 2)$  and $k_{\bar{q}}=(k-p / 2)$, respectively.
This BS amplitude can be written in terms of its Dirac operator structure and scalar functions:
\begin{eqnarray}
\Psi_{\pi}(k , p)& = & \gamma_{5}\, \chi_{1}(k, p)+\not k_{q} 
\gamma_{5}\, \chi_{2}(k, p)
\nonumber \\ 
& + &
\gamma_{5} k_{\bar{q}}\, \chi_{3}(k, p)+\not k_{q}
\gamma_{5} k_{\bar{q}}\, \chi_{4}(k, p)\, .
\end{eqnarray}
We aim to obtain the NIR weight functions of each scalar function $ \chi_{i}(k, p)$ within the
present chosen analytical model for the BS amplitude. For this purpose,
we introduce the useful identity given below:
\begin{small}
\begin{eqnarray}
	\label{ident}
\frac{1}{[(k+\frac{p}{2})^2-\mu^{\prime 2}+\imath\epsilon][k^2-\lambda^2+\imath\epsilon]
[(k-\frac{p}{2})^2-\mu^2+\imath\epsilon]} 
& = & \nonumber \\
\int_{0}^{\infty}  d\gamma\int_{-1}^1dz\frac{{F}(\gamma,z~;\mu^{\prime },\mu)}
{\left[k^2+z\, k \cdot P+\gamma+\imath\epsilon\right]^3}, \ \ \ 
\end{eqnarray}
\end{small}
where 
$$
 {F}(\gamma,z~;\mu^{\prime },\mu)=
\frac{2~\theta(1+z-2\alpha)~\theta(\alpha-z)~\theta(1-\alpha)~ \theta(\alpha)}
{|2\lambda^2+M^2/4-\mu^{\prime^2}-\mu^{2}|},
$$ 
and
$$
\alpha(\gamma,z~;\mu^{\prime },\mu)=
\frac{\gamma-z(\mu^{2}-\lambda^2-M^2/4)+\lambda^2}{2\lambda^2+M^2/4-\mu^{2}-\mu^{\prime^2}}.
$$
We can identify the four scalar functions of our model as:
\begin{eqnarray}
&&\chi_1(k, p)=- B(k_q^2)~ \frac{m^3\mathcal{N}}{k^2-\lambda^2+\imath\epsilon}~
  B(k_{\overline q}^2)\, ,\nonumber \\
&& \chi_2(k, p)=    - A(k_q^2) \frac{m^3\mathcal{N}}{k^2-\lambda^2+\imath\epsilon}~
              B(k_{\overline q}^2)\, ,
\nonumber\\
&& \chi_3(k,p)= -B(k_q^2)~ \frac{m^3\mathcal{N}}{k^2-\lambda^2+\imath\epsilon}~
  A(k_{\overline q}^2)\, ,
 \nonumber\\
 && \chi_4(k, p)=   -A(k_q^2)  \frac{m^3\mathcal{N}}{k^2-\lambda^2+\imath\epsilon}
A(k_{\overline q}^2)\, .
     \label{sistema0}
 \end{eqnarray}
In terms of the spectral representation of the dressed quark propagator the scalar amplitudes are
%%
%\begin{small}
 \begin{eqnarray}
\chi_{i}(k; p)  =  -\int_{0}^{\infty} d \mu^{\prime 2} 
\frac{\rho_{x_{i}}\left(\mu^{\prime 2}\right)}
{\left[(k+p / 2)^{2}-\mu^{\prime 2}+\imath \epsilon\right]} \quad
\nonumber \\
 \times \frac{\mathcal{N} \, m^{3}}
{\left[k^{2}-\lambda^{2}+\imath \epsilon\right]} 
\int_{0}^{\infty} d \mu^{2}
\frac{\rho_{y_{i}}\left(\mu^{2}\right)}
{\left[(k-p / 2)^{2}-\mu^{2}+v \epsilon\right]},\ \ \ \ 
\end{eqnarray}
%\end{small}
with the following  convention 
$\left(x_{1}, y_{1}\right) \equiv(B, B),\, \left(x_{2}, y_{2}\right) \equiv(A, B),\,
\left(x_{3}, y_{3}\right)
\equiv(B, A),$ and  $\left(x_{4}, y_{4}\right) \equiv(A, A)$.

Using the integral relation from Eq.~\eqref{ident}, we have that: 
 \begin{eqnarray}\label{eq:chi-int}
\chi_{i}(k, p) =  -\mathcal{N} m^{3}
\int_0^\infty
d \gamma \int_{-1}^{1} d z \int_{0}^{\infty} d \mu^{ ' 2} \int_{0}^{\infty} d \mu^{2}
\nonumber \\ 
\times \,
 \rho_{x_{i}} 
\left(\mu^{\prime 2}\right) \rho_{y_{i}}\,\left(\mu^{2}\right)
\frac{{F}\left(\gamma, z; \mu^{\prime}, \mu\right)}
{\left[k^{2}+z\, k \cdot p-\gamma+i \epsilon\right]^{3}}.
 \ \ \ \
\end{eqnarray}

A close inspection of Eq.~ \eqref{eq:chi-int} allows one to write the scalar amplitudes 
in terms of the Nakanishi integral representation,
$$
\chi_{i}(k, p)=\int_{-1}^{1} d z
\int_{0}^{\infty} d \gamma
\frac{G_{i}\left(\gamma, z\right)}
{\left[k^{2}+z\, k \cdot p-\gamma+i \epsilon\right]^{3}}\, ,
$$
where the weight functions are:
\begin{eqnarray}\label{eq:Gi}
G_{i}\left(\gamma, z\right)= 
\sum_{j=1}^{3} \sum_{k=1}^{3} C_{i;jk}\,\,{ F}\left(\gamma, z; m_{j}, m_{k}
\right)\, ,  
\end{eqnarray}
with the coefficients given by:
\begin{eqnarray}
C_{1;jk}& = & -\mathcal{N} m^{3} 
\left(E_{j}+m_{0} D_{j}\right)
\left(E_{k}+m_{0} D_{k}\right)\, , \nonumber  \\
C_{2;jk}& = & -\mathcal{N} m^{3}  D_{j}\left(E_{k}+m_{0} D_{k}\right) \, ,
\nonumber \\
C_{3;jk} & = & -\mathcal{N} m^{3}
\left(E_{j}+m_{0} D_{j}\right) 
D_{k}\, ,
\nonumber \\
C_{4;jk} & = & -\mathcal{N} m^{3}
 D_{j} D_{k} \, .
\label{gequations}
\end{eqnarray}
 Taking into account the properties under the exchange of indices of the coefficients
 above  and the explicit form of the NIR, 
we  have the following symmetry 
properties for the scalar amplitudes:
\begin{eqnarray}
&&\chi_{1}(k, p)  =  \chi_{1}(-k, p) ,\,\,
\chi_{2}(k, p)=  \chi_{3}(-k, p)\, , 
\nonumber \\
&&\chi_{4}(k, p)=\chi_{4}(-k, p)\, ,
\end{eqnarray}
which of course are consistent with the ones  easily derived from Eq.~\eqref{sistema0} 
with the explicit form of these amplitudes. These symmetries properties are also associated
with the even character in $z$ for
\begin{eqnarray}
G_1(\gamma,z) & = & G_1(\gamma,-z)~, \nonumber \\ 
& &   \quad\text{and}\quad G_4(\gamma,z)=G_4(\gamma,-z). 
\end{eqnarray}
The weight functions $G_2$ and $G_3$ in Eq.~\eqref{eq:Gi} are neither even or odd in $z$. 
However, due to  the symmetry property of the function $F\left(\gamma, z; m_{j}, m_{k}
\right)=F\left(\gamma,- z; m_{k}, m_{j}
\right)$ and $C_{2;jk}=C_{3,kj}$, they are related by $G_2(\gamma,z)=G_3(\gamma,-z)$.
Therefore, we chose to study  combinations of them, namely, 
$G_3(\gamma,z)+G_2(\gamma,z)$ and $G_3(\gamma,z)-G_2(\gamma,z)$,
which are even and odd in $z$, respectively.

%%  \end{multicols}

\begin{widetext}

\begin{figure}
\vspace{0.6cm}
    \centering
    \subfigure[]{\includegraphics[width=0.48\textwidth]{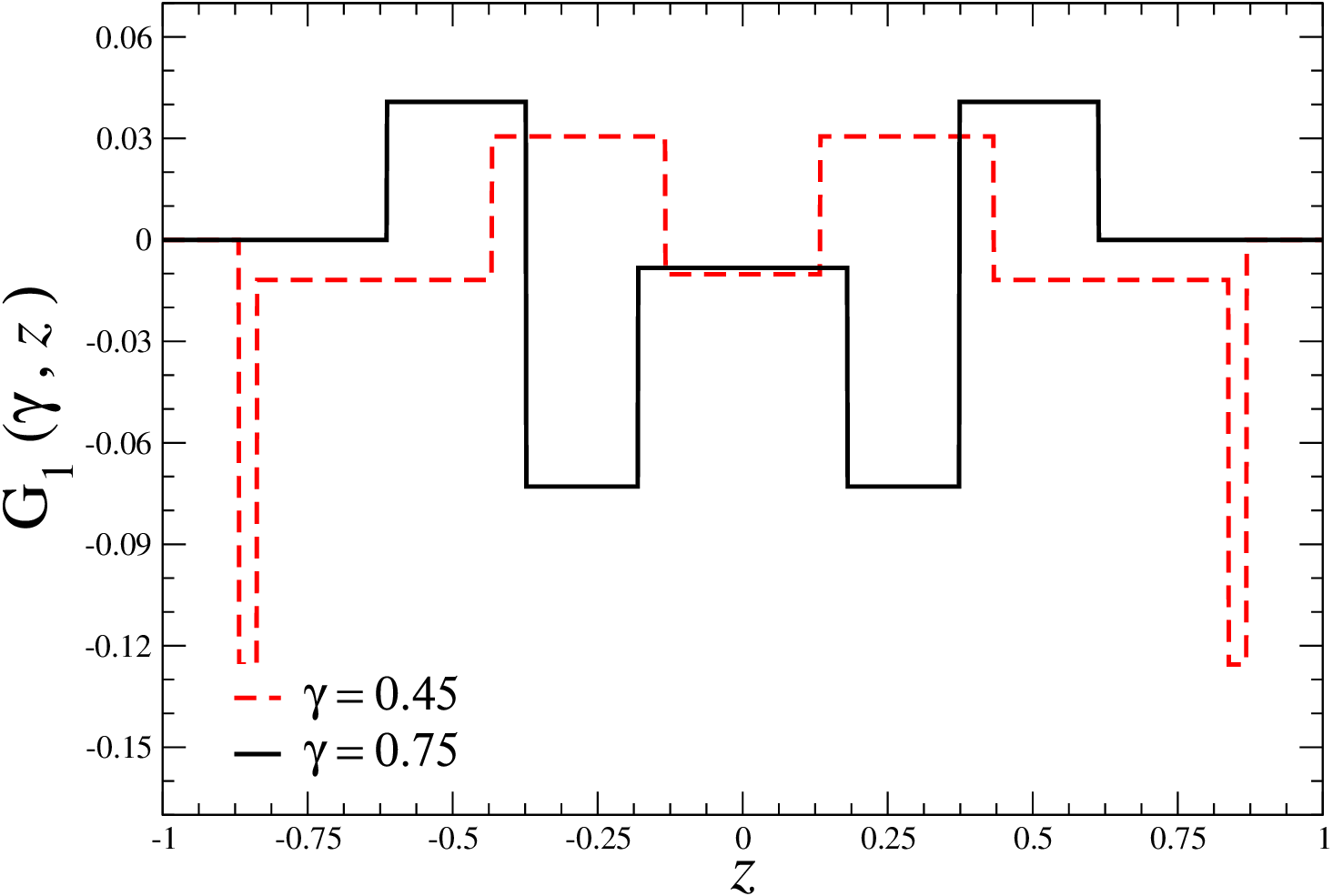}} 
    \subfigure[]{\includegraphics[width=0.48\textwidth]{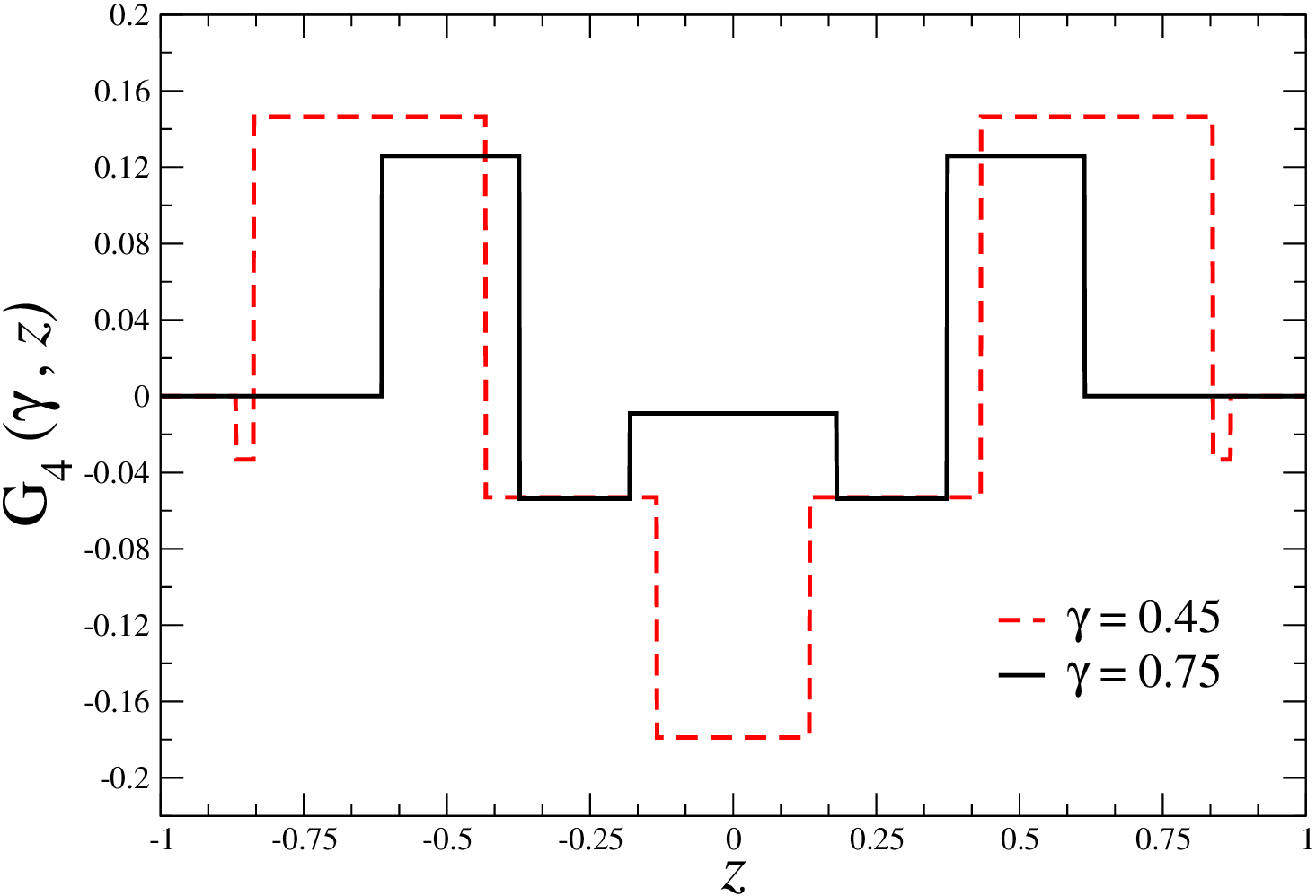}} 
    \caption{Left panel:~$G_1$ weight function dependence with $z$ for 
    $\gamma=0.45~\mathrm{GeV}^{2}$ (dashed line) and $0.75~\mathrm{GeV}^{2}$ (solid line).
    Right panel: $G_{4}(\gamma, z)$ as a function of $z$ with 
    $\gamma=0.45~\mathrm{GeV}^{2}$ (dashed line) and $0.75~\mathrm{GeV}^{2}$ (solid line). 
    The arbitrary value of $\mathcal{N}=100$ is used.}
    \label{fig12}
\end{figure}

\end{widetext}

%% \begin{multicols}{2}

\begin{widetext}
 
\begin{figure}
\vspace{0.5cm}
    \centering
   \subfigure[]{\includegraphics[width=0.48\textwidth]{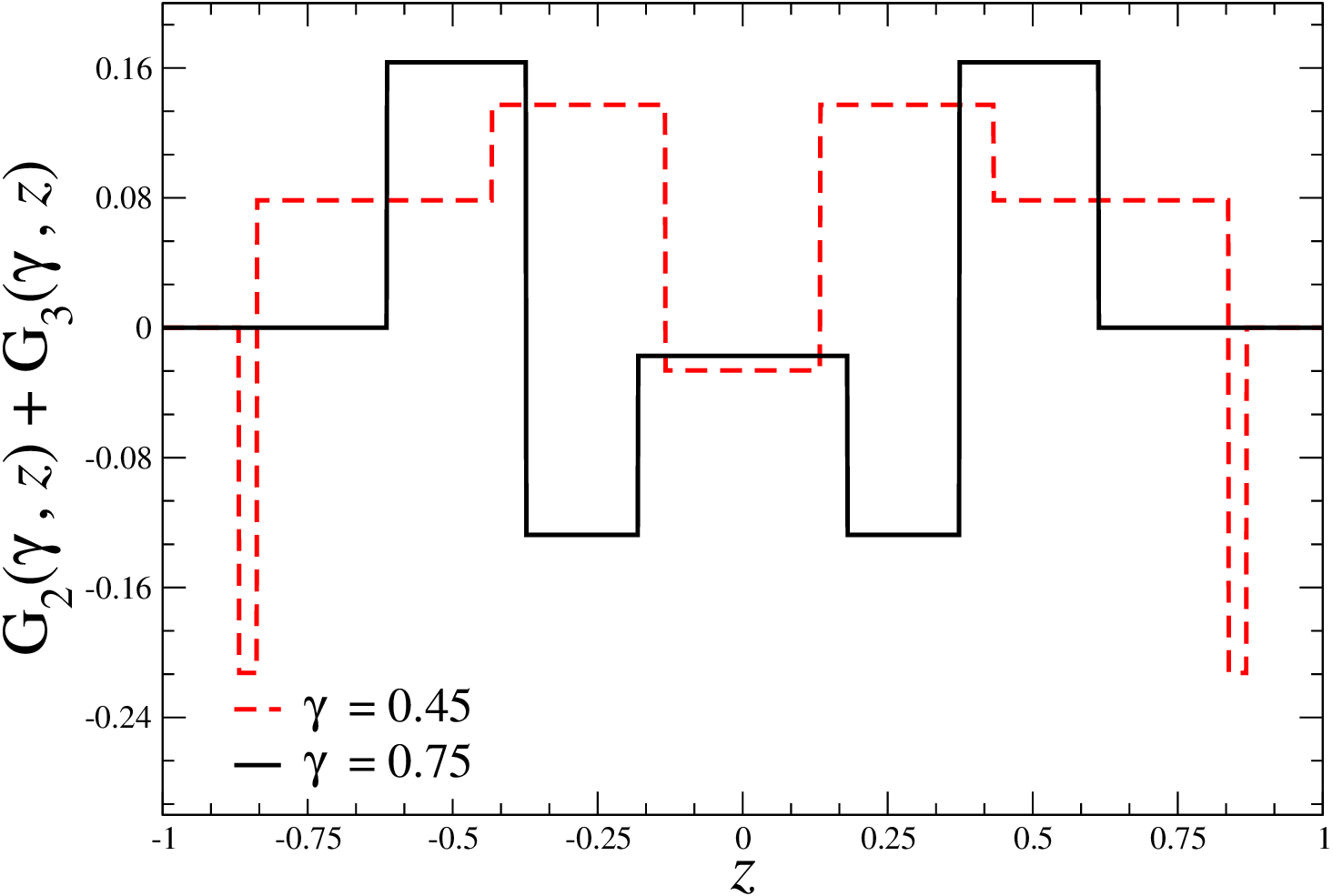}}
    \subfigure[]{\includegraphics[width=0.48\textwidth]{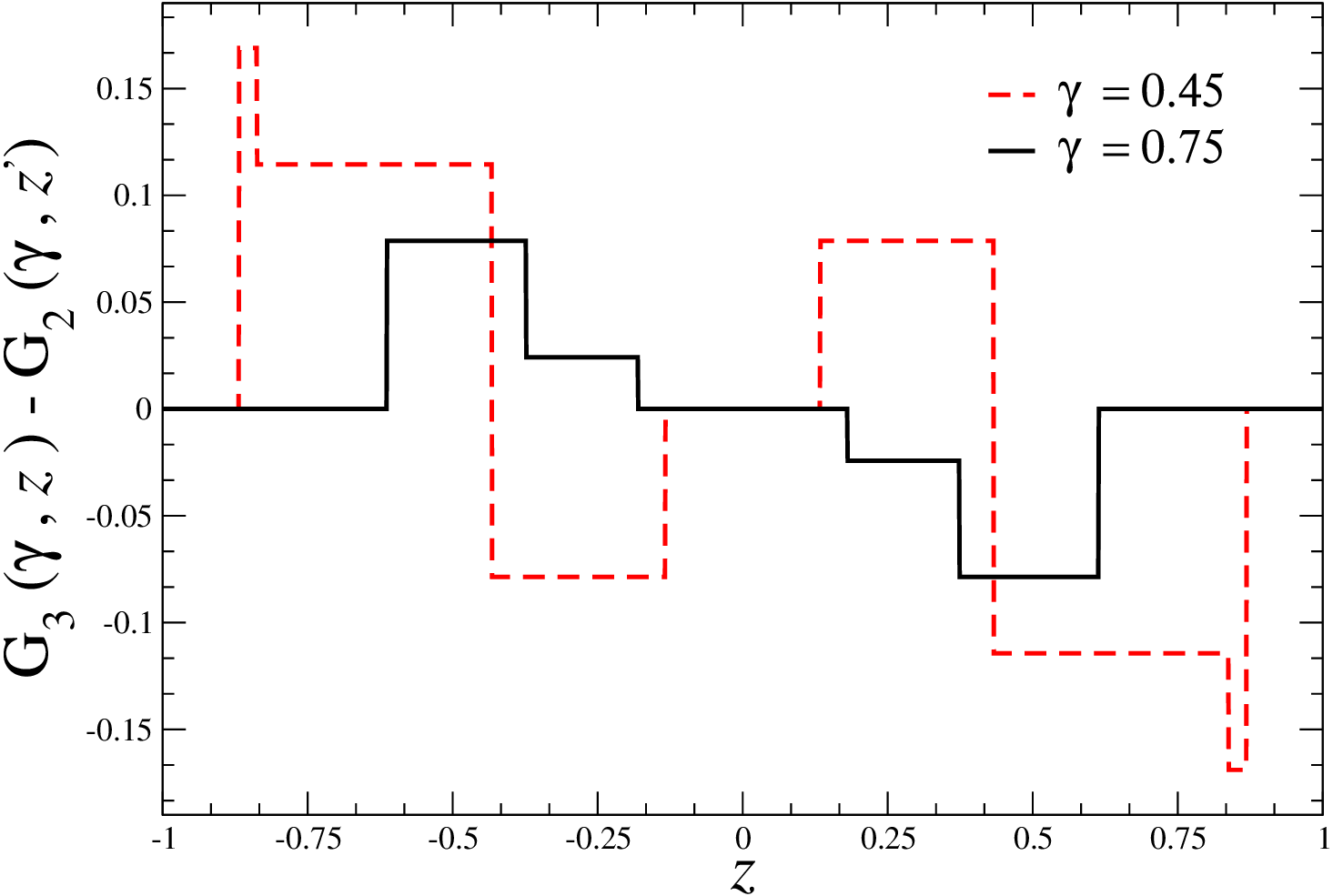}}
    \caption{Left panel:  $G_{2}~\left(\gamma, z\right)+G_{3}
    \left(\gamma, z\right)$
    as a function of $z$ for $\gamma=0.45 \ \mathrm{GeV}^{2}$ (dashed line) and 
    $0.75 \ \mathrm{GeV}^{2}$ (solid line). 
    Right panel: $G_{3}\left(\gamma, z\right)-G_{2}\left(\gamma, z\right)$ 
    as a function of $z$ for $\gamma=0.45~\mathrm{GeV}^{2}$ (dashed line) 
    and $0.75~\mathrm{GeV}^{2}$ (solid line). The arbitrary value of $\mathcal{N}=100$ is used.
        }
    \label{fig23}
\end{figure}

\end{widetext}

After the formal developments done so far, in what follows we present the numerical results for the four  
Nakanishi weight functions. For our purpose we study the dependence on $z$ of $G_i(\gamma,z)$ for 
$\gamma$ values of 0.45 and 0.75 GeV$^2$, which are within the scale of the mass poles of the dressed 
quark propagator and running mass function. The results are presented in figures \ref{fig12} 
and \ref{fig23}. The teeth-like structure of the weight functions are due to the overlap between
the theta functions present in the function $F(\gamma,z;\mu,mu')$, which are computed over the masses 
of the quark propagator poles, weighted by the coefficients $C_{i;jk}$ from Eq.~\eqref{gequations}, 
containing the residue of the functions $A(k^2)$ and $B(k^2)$ in the propagator. The 
different signs in the residue factors $E_i$ and $D_i$, which come with $C_{i;jk}$ are reflected 
in the jumping of the signs of $G_i$ when $z$ is varied, such behavior would be softened if smooth 
spectral functions associated with the quark propagator are in place, however if the positivity 
relations are to be violated an oscillating pattern should be expected for the Nakanishi weigth functions.

We observe in figures~\ref{fig12} and \ref{fig23} that all  $G_i(\gamma,z=\pm 1)$ vanish due to the property 
of $F(\gamma,z=\pm 1;\mu,\mu')=0$, which is essential to ensure that the pion valence light-front wave function
has the correct support in the longitudinal momentum fraction, vanishing at the end-points. The $G_i$ are quite 
sensitive to the variation of $\gamma$ from 0.45 to 0.75 GeV$^2$, which reflects the relevance of the infrared physics
of QCD to form the pion bound state, and responsible to give mass to the dressed quarks from the DCSB mechanism.
Essentially, the observed symmetry properties of $G_i$ with $z$ can be traced back to
the charge conjugation symmetry by the exchange of the quark and antiquark in the pion, as in our model
the $u$ and $d$ quarks are identical with respect to their self-energies.

Finally, we should mention that the four weight functions analyzed in this contribution  can
be used to describe the scalar functions associated with the decomposition of the
Bethe-Salpeter amplitude in the usual orthogonal basis of Dirac operators
(see e.g. \cite{dePaula:2016oct,dePaula:2017ikc}), and this will be covered 
in a future work, as well as the pion valence wave function \cite{Ydrefors:2021dwa} 
and momentum distributions\cite{dePaula:2020qna}, which can be written
in terms of the Nakanishi weight functions provided here.

{\it Acknowledgements:}~
This work was supported in part by CAPES under Grant No. 88881.309870/2018-01 (WdP), and
by the Conselho Nacional de Desenvolvimento 
Cient\'{i}fico e Tecnol\'{o}gico (CNPq), Grant No.~308486/2015-3 (TF),
Process No.~307131/2020-3 (JPBCM),
Grants No. 438562/2018-6 and No. 313236/2018-6 (WdP) and Funda\c{c}\~{a}o de Amparo \`{a}
Pesquisa do Estado de S\~{a}o Paulo (FAPESP), Process No. 2019/02923-5 (JPBCM),  
and was also part of the projects, Instituto Nacional de Ci\^{e}ncia e
Tecnologia -- Nuclear Physics and Applications (INCT-FNA), Brazil,
Process No.~464898/2014-5, and FAPESP Tem\'{a}tico, Brazil, Process,
the thematic projects, No. 2013/26258-4 and No. 2017/05660-0. 
%%% \onecolumngrid

%%% \clearpage  \newpage 

%%   \medline

%% \begin{multicols}{2}

%%   \end{multicols}

\end{document}